\documentstyle[aps]{revtex}
\begin{document}
%\topmargin=0in
%\headheight=0in
%\headsep=0in
%\oddsidemargin=7.2pt
%\evensidemargin=7.2pt
%\footheight=1in
%\marginparwidth=0in
%\marginparsep=0in
%\textheight=235mm
%\textwidth=160mm

%\documentstyle[12pt]{article}
%\setlength{\textwidth}{6in}
%\setlength{\textheight}{8.0in}
%\setlength{\parskip}{0.05in}
%\flushbottom
%\setlength{\baselineskip}{15.2pt}
%\setlength{\baselineskip}{24.2pt}

%
% section.equation numbering style --- off
%\renewcommand{\theequation}{\thesection.\arabic{equation}}

\newcommand\ie {{\it i.e. }}
\newcommand\eg {{\it e.g. }}
\newcommand\etc{{\it etc. }}
\newcommand\cf {{\it cf.  }}
\newcommand\etal {{\it et al. }}
\newcommand{\be}{\begin{eqnarray}}
\newcommand{\ee}{\end{eqnarray}}
\newcommand\Jpsi{{J/\psi}}
\newcommand\M{M_{Q \overline Q}}
\newcommand\mpmm{{\mu^+ \mu^-}}
\newcommand{\jp}{$ J/ \psi $}
\newcommand{\pp}{$ \psi^{ \prime} $}
\newcommand{\ppp}{$ \psi^{ \prime \prime } $}
\newcommand{\dd}[2]{$ #1 \overline #2 $}
\newcommand\noi {\noindent}
\draft
\preprint{LBL-36754 \  SLAC-PUB-95-6753}
\title
{Intrinsic Charm Contribution to Double Quarkonium
Hadroproduction$^\star$} \footnotetext{$^\star$ This work was
supported in part by the Director, Office of Energy Research,
Division of Nuclear Physics of the Office of High Energy and Nuclear
Physics of the U. S. Department of Energy under Contract Numbers
DE-AC03-76SF0098 and DE-AC03-76SF00515.}

\author{R. Vogt}
\address
{Nuclear Science Division,
Lawrence Berkeley Laboratory, Berkeley, CA 94720,
USA \\
and \\
Institute for Nuclear Theory,
University of Washington, Seattle, WA 98195,
USA}

\author{S. J. Brodsky}
\address
{Stanford Linear Accelerator Center,
Stanford University, Stanford, CA 94309,
USA}
\maketitle

\begin{abstract}
Double $J/\psi$ production has been observed by the NA3
collaboration in $\pi N$ and $p N$ collisions
with a cross section of the order of 20-30 pb. The
$\psi \psi$ pairs measured in $\pi^- $ nucleus interactions at 150
and 280 GeV$/c$ are observed to carry an anomalously large fraction
of the projectile momentum in the laboratory frame, $x_{\psi \psi}
\geq 0.6$ at 150 GeV$/c$ and $\geq 0.4$ at 280 GeV$/c$. We postulate
that these forward $\psi \psi$ pairs are created by the
materialization of Fock states in the projectile containing two
pairs of intrinsic $c \overline c$ quarks. We calculate the overlap
of the charmonium states with the $|\overline u d c \overline c c
\overline c \rangle$ Fock state as described by the intrinsic charm
model and find that the $\pi^- N \rightarrow \psi \psi$ longitudinal
momentum and invariant mass distributions are both well reproduced. We
also discuss double $J/\psi$ production in $pN$ interactions and the
implications for other heavy quarkonium production channels in QCD.
\end{abstract}
\newpage

\vspace{1.0cm}
\begin{center}
{\bf Introduction}
\end{center}

It is quite rare for two charmonium states to be produced in the
same hadronic collision. However, the NA3 collaboration has measured
a double $J/\psi$  production  rate significantly above background
in multi-muon events with $\pi^-$ beams at laboratory momentum
150 and 280 GeV/c
\cite{Badpi} and a 400 GeV/c proton beam \cite{Badp}. The integrated
$\pi^- N \rightarrow \psi \psi X$ production cross section,
$\sigma_{\psi \psi}$, is $18 \pm 8$ pb at 150 GeV/c and $30 \pm 10$ pb
at 280 GeV/c, and the $pN \rightarrow \psi \psi X$ cross section is $27
\pm 10$ pb.  The relative double to single rate, $\sigma_{\psi
\psi}/\sigma_\psi$, is $(3 \pm 1) \times 10^{-4}$ for pion-induced
production where $\sigma_\psi$ is the integrated single $\psi$
production cross section.

A particularly surprising feature of the NA3 $\pi^-N \rightarrow
\psi\psi X$ events is that the laboratory fraction of the projectile
momentum carried by the $\psi \psi$ pair is always very large,
$x_{\psi \psi} \geq 0.6$ at 150 GeV/c and $x_{\psi \psi} \geq 0.4$ at
280 GeV/c.  In some events, nearly all of the projectile momentum is
carried by the $\psi \psi$ system.
%Additionally, single $\psi$'s in the pairs have $x_\psi > 0.15$.
In contrast, perturbative $ g g$
and $q \overline q$ fusion processes are expected to produce central
$\psi \psi$ pairs, centered around the mean value, $\langle x_{\psi
\psi} \rangle \approx$ 0.4-0.5, in the laboratory
\cite{BHK,ES,HM,Russ}.

The average invariant mass of the pair, $\langle M_{\psi \psi} \rangle = 7.4$
GeV, is well
above the $2m_\psi$ threshold.  In fact, all the events have $M_{\psi \psi} >
6.7$ GeV.  The average transverse
momentum of the pair is quite small, $p_{T,\psi \psi} = 0.9 \pm 0.1$
GeV, suggesting that $\psi \psi$ pair production is highly
correlated \cite{Badpi,Badp}. The proton events have a somewhat
lower invariant mass, $\langle M_{\psi \psi} \rangle \approx 6.8$
GeV.  The  $x_{\psi \psi}$ distribution for the $pN\rightarrow \psi\psi
X$ events has not been reported.

There have been attempts to explain the NA3 data within
conventional leading-twist QCD. Charmonium pairs can
be produced by a variety of QCD processes including $B \overline
B$ production and decay, $B \overline B \rightarrow \psi \psi X$
\cite{BHK} and ${\cal O}(\alpha_s^4)$ $\psi \psi$ production via $gg$
fusion and $q \overline q$ annihilation \cite{ES,HM,Russ}.
Li and Liu have also considered the possibility  that a
$2^{++} c \overline c c \overline c$ resonance is
produced which then decays into correlated $\psi
\psi$ pairs \cite{LL}. All of these models predict centrally
produced $\psi \psi$ pairs \cite{BHK,ES,HM,Russ}, in contradiction
to the $\pi^-$ data. In addition, the predicted magnitude of
$\sigma_{\psi \psi}$ is too small by a factor of 3-5.  If these
models are updated using recent branching ratios and current
scale-dependent parton distributions, the predicted
leading twist cross sections are
further reduced\cite{RV},
suggesting that an additional mechanism is needed
to produce fast $\psi \psi$ pairs.

We also note that strong
discrepancies between conventional QCD predictions
and experiment have recently
been observed for $\psi$, $\psi^\prime$, and $\Upsilon$ production
at large $p_T$ in high energy $p \overline p$ collisions
at the Tevatron \cite{Tev}.  Braaten and Fleming\cite{BrF}
have suggested that
this surplus of
charmonium production is due to
the enhanced fragmentation of gluon jets coupling to the octet
$c \overline c$ components
in higher Fock states $|c \overline c g g \rangle$ of the charmonium
wavefunction.  However, this explanation does not explain the
NA3 $\psi \psi$ fixed target data
since the color-octet mechanism still emphasizes central production.

In this paper we shall explore an alternative higher-twist
production mechanism in QCD
where forward single and double $\psi$'s are created through the
materialization of intrinsic $c \overline c$ Fock components of the
pion or proton projectile \cite{intc}.  In such states the
heavy constituents tend to
carry the majority of the projectile momentum since this
minimizes the off-shell energy of the wavefunction.
We find that the $\psi \psi$ momentum and mass
distributions are reasonably well described by the overlap of the
charmonium states with the  $|\overline u d c \overline c c
\overline c \rangle$ Fock configuration postulated in the intrinsic
heavy quark model.

The QCD wavefunction of a hadron can be represented as a
superposition of quark and gluon Fock states. For example, at fixed
light-cone time, $\tau= t + z/c$, the $\pi^-$ wavefunction can be
expanded as a sum over the complete basis of free quark and gluon
states: $\vert \Psi_{\pi^-}\rangle = \sum_m \vert m \rangle \,
\psi_{m/\pi^-}(x_i, k_{T,i}, \lambda_i)$ where the color-singlet
states, $\vert m \rangle$, represent the Fock components $\vert
\overline  u  d \rangle$, $\vert \overline u d g \rangle$, $\vert
\overline u d Q \overline Q \rangle$, {\it etc}. The boost-invariant
light-cone wavefunctions, $\psi_{m/\pi^-}(x_i, k_{T,i}, \lambda_i)$,
needed to compute probability distributions, are functions of the
relative momentum coordinates $x_i = k_i^+/P^+$ and $k_{T,i}$.
Momentum conservation demands $\sum_{i=1}^n x_i = 1$ and
$\sum_{i=1}^n \vec{k}_{T,i}=0$, where $n$ is the number of partons
in a Fock state $\vert m \rangle$.  When the projectile scatters in
the target, the coherence of the Fock components is broken, its
fluctuations can hadronize, forming new hadronic systems from the
fluctuations \cite{BHMT}. For example, intrinsic $c \overline c$
fluctuations can be liberated provided the system is probed during
the characteristic time, $\Delta t = 2p_{\rm lab}/M^2_{c \overline
c}$, that such fluctuations exist.

Microscopically, the intrinsic heavy quark Fock component in the
$\pi^-$ wavefunction, $|\overline u d Q \overline Q \rangle$, is
generated by virtual interactions such as $g g \rightarrow Q
\overline Q$ where the gluons couple to two or more projectile
valence quarks. The probability for $Q \overline Q$ fluctuations to
exist in a light hadron thus scales as $\alpha_s^2(m_Q^2)/m_Q^2$
relative to leading-twist production \cite{VB}. Therefore, this
contribution is higher twist,
%RV order (1/m_Q^2) correct?
suppressed by ${\cal O}(1/m_Q^2)$ compared to sea quark
contributions generated by gluon splitting.   In $(1+1)$ QCD the
Fock state representation of each hadron can be computed explicitly,
including its intrinsic $Q \overline Q$ configurations, by
diagonalizing a discrete form of the light-cone Hamiltonian
\cite{HBP}.

In general, the dominant Fock state configurations are not far off
shell and thus have minimal invariant mass, $M^2 = \sum_i m_{T,
i}^2/ x_i$ where $m_{T, i} = \sqrt{k^2_{T,i}+m^2_i}$ is the
transverse mass of the $i^{\rm th}$ particle in the configuration.
Intrinsic $Q \overline Q$ Fock components with minimum invariant
mass correspond to configurations with equal rapidity constituents.
Thus, unlike sea quarks generated from a single parton, intrinsic
heavy quarks tend to carry a larger fraction of the parent momentum
than the light quarks \cite{intc}. In fact, if the intrinsic $Q
\overline Q$ coalesces into a quarkonium state, the momentum of the
two heavy quarks is combined so that the quarkonium state will carry
a significant fraction of the projectile momentum. It was shown that
large $x_F$ virtual $c\overline c$ or lepton pairs can be liberated
by a relatively soft interaction with a light quark component of the
projectile \cite{BHMT}. For soft interactions at momentum scale
$\mu$, the intrinsic heavy quark cross section is suppressed by a
resolving factor $\propto \mu^2/m^2_Q$ \cite{VB}.

There is substantial circumstantial evidence for
the existence of intrinsic $c
\overline c$ states in light hadrons. For example, the charm
structure function of the proton measured by EMC is significantly
larger than predicted by photon-gluon fusion at large $x_{Bj}$
\cite{EMCic}. Leading charm production in $\pi N$ and hyperon-$N$
collisions also requires a charm source beyond leading twist
\cite{VB,769}.  The NA3 experiment has also shown that the
single $J/\psi$ cross section at large $x_F$ is greater than
expected from $gg$ and $q \overline q$ production \cite{Bad}.
Additionally, intrinsic charm may account for the anomalous
longitudinal polarization of the $J/\psi$
at large $x_F$ \cite{Van} seen in $\pi N \to J/\psi X $ interactions.

Over a sufficiently short time, the pion can contain Fock states of
arbitrary complexity. For example, two intrinsic $c \overline c$
pairs may appear simultaneously in the quantum fluctuations of
the projectile wavefunction and then,
freed in an energetic interaction, coalesce to form a pair of
$\psi$'s. We shall estimate the creation probability of $|n_V c
\overline c c \overline c \rangle$ Fock states, where $n_V = d
\overline u$ for $\pi^-$ and $n_V = uud$ for proton projectiles, assuming
that all of the double $J/\psi$ events arise from these
configurations.  We then examine the $x_{\psi \psi}$ and invariant
mass distributions of the $\psi \psi$ pairs and the $x_\psi$
distribution for the single $\psi$'s arising from these Fock states.

\vspace{1.0cm}
\begin{center}
{\bf Intrinsic Charm Fock States}
\end{center}

The probability distribution for a general $n$--particle intrinsic
$c \overline c$ Fock state as a function of $x$ and $\vec{k}_T$ is
written as
\be
\frac{dP_{\rm ic}}{\prod_{i=1}^n dx_i d^2k_{T, i}} =
N_n \alpha_s^4(M_{c \overline c})\ \frac{\delta(\sum_{i=1}^n \vec
k_{T, i}) \delta(1-\sum_{i=1}^n x_i)} {(m_h^2 - \sum_{i=1}^n (m_{T,
i}^2/x_i) )^2} \, \, ,
\ee
where $N_n$ normalizes the Fock state
probability.
In the model,  the vertex function
in the intrinsic charm wavefunction is assumed to be
relatively slowly varying; the particle distributions are then
controlled by the light-cone energy denominator and  phase space.
This form for the higher Fock wavefunctions
generalizes for an
arbitrary number of light and heavy quark components.
The Fock states containing charmed quarks can be materialized
by a soft collision in the target which brings the state
on shell. The distribution of produced open and hidden charm states
will reflect the underlying shape of the Fock state wavefunction.

The invariant mass of a $c \overline c$ pair, $M_{c \overline c}$, from such a
Fock state is
\be
\frac{dP_{\rm ic}}
{dM^2_{c \overline c}} & = & \int \prod_{i=1}^n dx_i
d^2k_{T, i} \ \frac{dx_{c \overline c}}{x_{c \overline c}}\
d^2k_{T, c \overline c}
\ \frac{dP_{\rm ic}}{\prod_{i=1}^n dx_i d^2k_{T, i}}\
\delta(x_{c \overline c} - x_c - x_{\overline c})
\\ \nonumber
&  & \mbox{} \times \delta(\vec k_{T, c} + \vec k_{T, \overline c} - \vec
k_{T, c \overline c}) \,
\delta \left( \frac{M^2_{T, c \overline c}}{x_{c \overline c}} -
\frac{m_{T, c}^2}{x_c} -
\frac{m_{T, \overline c}^2}{x_{\overline c}} \right)
\, \, ,
\ee
where $n=4$ and 5 is the number of partons in the lowest lying meson
and baryon intrinsic $c \overline c$ Fock states. The probability to
produce a $J/\psi$ from an intrinsic $c \overline c$ state is
proportional to the fraction of intrinsic $c \overline c$ production
below the $D \overline D$ threshold. The fraction of $c \overline c$
pairs with $2m_c < M_{c \overline c} < 2m_D$ is
\be
f_{c \overline c/h} = \int_{4 m^2_c}^{4
m_D^2} dM^2_{c \overline c} \ \frac{dP_{\rm ic}}{dM^2_{c \overline c}}
\ \Bigg/ \int_{4 m^2_c}^s %{\infty}
dM^2_{c \overline c} \ \frac{dP_{\rm ic}}{dM^2_{c \overline c}}
\, \, .
\ee
The ratio $f_{c \overline c/\pi}$ is approximately 15\%\ larger than
$f_{c \overline c/p}$ for $1.2 < m_c < 1.8$ GeV.  However, not all
$c \overline c$'s produced below the $D \overline D$ threshold will
produce a final-state $J/\psi$. We include two suppression factors to
estimate $J/\psi$ production, one reflecting
the number of quarkonium
channels available
with $M_{c \overline c} < 2m_D$ and one for the $c$ and
$\overline c$ to coalesce with each other rather than combine with
valence quarks to produce open charm states.
The ``channel"
suppression factor, $s_c \approx 0.3$, is estimated from direct and
indirect $J/\psi$ production, including $\chi_1$ and $\chi_2$
radiative and $\psi^\prime$ hadronic decays. The combinatoric
``flavor" suppression factor, $s_f$, is $1/2$ for a $|\overline u d
c \overline c \rangle$ state and $1/4$ for a $|uudc \overline c
\rangle$ state.
In Fig.\ 1 we show the predicted fraction of $\psi$'s produced
from intrinsic $c \overline c$ pairs,
\be
f_{\psi/h} = s_c s_f f_{c \overline c/h} \, \, ,
\ee
as a function of $m_c$.  We take $m_c =
1.5$ GeV, suggesting $f_{\psi/\pi} \approx 0.03$ and $f_{\psi/p}
\approx 0.014$.

The nuclear dependence arising from the manifestation of intrinsic charm
is expected to be $\sigma_A \approx \sigma_N A^{2/3}$, characteristic of
soft interactions.  Indeed, in their single $J/\psi$ measurements \cite{Bad},
the NA3 collaboration defined a separate ``diffractive" cross section at large
$x_F$ scaling like $A^{0.77}$ for $\pi A$ and $A^{0.71}$ for $pA$ interactions.
We thus identify the diffractive cross section with $\sigma_{\rm ic}^\psi$,
as in our previous study \cite{VBH1}.
The intrinsic $c \overline c$ production
cross section from an $|n_V c \overline c \rangle$ configuration,
\be
\sigma_{\rm ic}(hN) = P_{\rm ic}
\sigma_{h p}^{\rm in} \ \frac{\mu^2}{4 \widehat{m}_c^2} \, \, ,
\ee
was estimated in Ref.\ \cite{VB} to be $\sigma_{\rm ic}(\pi N)
\approx$ 0.5 $\mu$b and $\sigma_{\rm ic}(pN) \approx$ 0.7 $\mu$b for a
beam momentum of 200 GeV.
At low $x_F$, a different, nearly linear, $A$ dependence was observed.
The ratio of diffractive to total $J/\psi$ production can then
be determined from the NA3 data \cite{Bad}.  The soft interaction
scale parameter \cite{chev}, $\mu^2 \sim 0.2$ GeV$^2$, is thus fixed by the
assumption that the diffractive fraction of the total production
cross section is the same for charmonium and charmed hadrons.

The diffractive $J/\psi$ cross section is $(18 \pm 3)$\%\ of the total
$\pi A \rightarrow J/\psi$ production cross section and $(29 \pm 6)$\%\
of the $pA \rightarrow J/\psi$ production \cite{Bad}.
The cross sections in the $J/\psi
\rightarrow \mu^+ \mu^-$ channel are $B\sigma_\psi(\pi^- N) = 6.5
\pm 0.6$ nb at 150 GeV and $B\sigma_\psi(p N) = 3.6 \pm 0.6$ nb at
200 GeV.
Removing the branching ratio \cite{Bad}, then implies $\sigma_{\rm
ic}^\psi(\pi^- N) = 15.7 \pm 3.0$ nb and $\sigma_{\rm ic}^\psi (pN)
= 14.3 \pm 3.3$ nb.  The proton cross section should be regarded as
an upper bound since the bulk of this cross section is at $x_F <
0.2$ where the intrinsic charm contribution is expected to be small.
Using the data for $x_F > 0.2$ suggests a lower bound, $\sigma_{\rm
ic}^\psi (pN) = 5.6 \pm 1.3$ nb.  We relate $\sigma_{\rm ic}^\psi$
to the intrinsic $c \overline c$ production cross section by
\be
\sigma_{\rm ic}^\psi(hN) = f_{\psi/h} \sigma_{\rm ic} (hN) \, \, .
\ee
Using the estimates of $\sigma_{\rm ic}
(\pi N)$ from Ref.\ \cite{VB} with the values of $f_{\psi/h}$ in Eq.\ (4),
we find $\sigma_{\rm
ic}^\psi (\pi^- N) = 15$ nb and $\sigma_{\rm ic}^\psi (pN) = 9.8$
nb, in agreement with the NA3 single $\psi$
data\footnote{We use $B(J/\psi \rightarrow \mu^+ \mu^-) \approx
7.4$\%\
for consistency with the NA3 measurements.  If $B(J/\psi \rightarrow
\mu^+ \mu^-) \approx 6$\%\ is used instead, the single cross sections
increase by 20\%.}.

\vspace{1.0cm}
\begin{center}
{\bf Double $J/\psi$ Production from Intrinsic Charm Fock States}
\end{center}

If one assumes that all of the NA3 double $J/\psi$ events arise from
intrinsic $|n_V c \overline c c \overline c \rangle$ Fock states, then the
required normalization for this state can be determined from
\be
\sigma_{\rm ic}^{\psi \psi} (hN) =
f_{\psi/h}^2 \ \frac{P_{\rm icc}}{P_{\rm ic}}\  \sigma_{\rm ic} (hN)
= f_{\psi/h} \ \frac{P_{\rm icc}}{P_{\rm ic}}\  \sigma_{\rm ic}^\psi (hN)
\ee
where $P_{\rm icc}$ is the probability to produce a pair of
intrinsic $c \overline c$ states in the projectile. Then
$\sigma_{\psi \psi} \equiv \sigma_{\rm ic}^{\psi \psi} (\pi^- N)
\approx 20$ pb \cite{Badpi}, requiring $P_{\rm icc} \approx 4.4\%\
P_{\rm ic}$. If $P_{\rm icc}$ is independent of the projectile, then
we predict $\sigma_{\rm ic}^{\psi \psi}(pN) \approx 6$ pb, which is
20\%\ of the measured cross section \cite{Badp}, but consistent with
our previous calculations \cite{VB}.  If, instead, the ratio
$\sigma_{\psi \psi}/\sigma_\psi$ \cite{Badpi} is assumed to be
independent of the projectile, the cross section increases to 14.6
pb, more compatible with the data and $P_{\rm icc} \approx 10.6\%\
P_{\rm ic}$. Finally, if we assume $\sigma_{\rm ic}^{\psi \psi} (p N) = 27$ pb,
as suggested by the data of Ref.\ \cite{Badpi}, then we require
$P_{\rm icc} \approx 20\%\ P_{\rm ic}$.  Note that the probability of
making a second intrinsic $c \overline c$ pair is relatively easy once one $c
\overline c$ is present in the Fock state.

The $\psi \psi$ invariant mass distributions measured by NA3
\cite{Badpi,Badp} are shown in the histograms of Fig.\ 2.
The pair
mass distribution predicted from the $|n_V c \overline c c \overline c
\rangle$ Fock state is
\be
\lefteqn{\frac{dP_{\rm icc}}{dM^2_{\psi
\psi}} = \int \prod_{i=1}^n dx_i d^2k_{T, i} \prod_{j=1}^2
\ \frac{dx_{\psi_j}}{x_{\psi_j}}\ dm_{\psi_j}^2 d^2k_{T, \psi_j}
\ \frac{dx_{\psi \psi}}{x_{\psi \psi}}\ d^2k_{T, \psi \psi}
\ \frac{dP_{\rm icc}}{\prod_{i=1}^n dx_i d^2k_{T, i}}} \\ \nonumber &
& \mbox{} \times \delta \left( \frac{m^2_{T, \psi_j}}{x_{\psi_j}} -
\frac{m_{T, c_j}^2}{x_{c_j}} - \frac{m_{T, \overline
c_j}^2}{x_{\overline c_j}} \right) \delta(\vec k_{T, c_j} + \vec
k_{T, \overline c_j} - \vec k_{T, \psi_j}) \delta(x_{\psi_j} -
x_{c_j} - x_{\overline c_j}) \\ \nonumber &  & \mbox{} \times\delta
\left( \frac{M^2_{T, \psi \psi}}{x_{\psi \psi}} - \frac{m_{T,
\psi_1}^2}{x_{\psi_1}} - \frac{m_{T, \psi_2}^2}{x_{\psi_2}} \right)
\delta(\vec k_{T, \psi_1} + \vec k_{T, \psi_2} - \vec k_{T, \psi
\psi}) \delta(x_{\psi \psi} - x_{\psi_1} - x_{\psi_2}) \, \, ,
\ee
where $n = n_V + 4$ and
\be
\frac{dP_{\rm icc}}{\prod_{i=1}^n dx_i
d^2k_{T, i}} = N_n [\alpha_s^4(M_{c \overline c})]^2
\ \frac{\delta(\sum_{i=1}^n \vec k_{T, i}) \delta(1-\sum_{i=1}^n
x_i)}{(m_h^2 - \sum_{i=1}^n (m_{T, i}^2/x_i) )^2} \, \, .
\ee
The delta functions insure conservation of momentum for both $J/\psi$
mesons and the $\psi \psi$ pair.  Here, $2m_c < m_\psi < 2m_D$ and
$m_c = 1.5$ GeV.  In Fig.\ 2 the predictions for the $\psi \psi$
pair mass distributions are compared with the $\pi^- N$ (a) and $pN$
(b) data and normalized to the data. Without any $k_T$ dependence,
the mass distribution is strongly peaked at threshold.  The $k_T$
dependence smears out the pair distribution, increasing $\langle
M_{\psi \psi} \rangle$ by $0.5 - 1.0$ GeV.  We find $\langle M_{\psi \psi}
\rangle \approx 7.7$ GeV for the pion beam and $\langle M_{\psi
\psi} \rangle \approx 7.4$ GeV for the proton beam.  The intrinsic
charm model is very successful in reproducing the strongly
correlated features of the data. The smaller value of $\langle
M_{\psi \psi} \rangle$ in $pN$ interactions also agrees with the
trend of the data \cite{Badpi,Badp}.

Although the
$\vec k_T$ dependence is needed to calculate the mass distributions,
it is sufficient to
use a mean value of $k_T^2$ to calculate the $x$ distributions since
the shapes of the longitudinal momentum distributions are
essentially independent of the transverse momentum.  Then the $x$ distributions
from an $|n_V c \overline c c \overline c \rangle$
state, following Eq.\ (1), can be written as
\be
\frac{dP_{\rm icc}}{dx_i \cdots dx_n} =  N_n [\alpha_s^2(M_{c
\overline c})]^2
\ \frac{\delta(1-\sum_{i=1}^n x_i)}{(m_h^2 - \sum_{i=1}^n
(\widehat{m}_i^2/x_i)
)^2} \, \, ,
\ee
where again $n = n_V +4$ and $\widehat{m}_i = \sqrt{m_i^2 + \langle
\vec k_{T, i}^2 \rangle}$ is the average transverse mass. Assuming
$\langle \vec k_{T, i}^2 \rangle$ is proportional to the square of
the constituent quark mass, we adopt the effective values
$\widehat{m}_c = 1.8$ GeV and $\widehat{m}_q = 0.45$ GeV, as in our
previous work \cite{VBH1,VBH2}. The $x_\psi$ distribution for a
single $J/\psi$ in a $|n_V c \overline c c \overline c \rangle$
state is
\be
\frac{dP_{\rm icc}}{dx_{\psi_1}} = \int \prod_{i=1}^n
dx_i dx_{\psi_2} \ \frac{dP_{\rm icc}}{dx_1 \ldots dx_n}\
\delta(x_{\psi_1} - x_{c_1} -x_{\overline c_1}) \delta(x_{\psi_2} -
x_{c_2} - x_{\overline c_2}) \, \, .
\ee
We find $\langle x_\psi \rangle = 0.36$ for the pion projectile and
$\langle x_\psi \rangle = 0.33$ for the proton. The number of single
$J/\psi$'s is twice the number of $\psi \psi$ pairs. The single
$J/\psi$ distributions from the $|n_V c \overline c c \overline c
\rangle$ state have a lower average $x_\psi$ than those from $|n_V c
\overline c \rangle$ Fock states, where $\langle x_\psi \rangle =
0.62$ for a pion and $\langle x_\psi \rangle = 0.51$ for a proton
\cite{Bad,VBH1}.  The pair distribution can be computed from
\be
\frac{dP_{\rm icc}}{dx_{\psi \psi}} = \int dx_{\psi_1} \ \frac{dP_{\rm
icc}}{dx_{\psi_1}}\ \delta(x_{\psi \psi} - x_{\psi_1} - x_{\psi_2}) \,
\, .
\ee
The intrinsic charm model predicts  $\langle x_{\psi \psi}
\rangle = 0.72$ for the pion and $\langle x_{\psi \psi} \rangle =
0.64$ for the proton. We compare the frame-independent calculation
to histograms of the combined 150 and 280 GeV data assuming $x_\psi =
p_{\rm lab}^\psi/p_{\rm beam}$ in Figure \ 3.
The $\psi \psi$ pair distributions
are shown
in Fig.\ 3(a) and 3(c) and the associated
the single $J/\psi$ distributions in pair events are
shown in Fig.\ 3(b) and 3(d).  Both are normalized to the data with
the single $J/\psi$ normalization twice that of the pair.

\vspace{1.0cm}
\begin{center}
{\bf Other Tests of the Intrinsic Heavy Quark Mechanism}
\end{center}

The intrinsic charm model provides a natural explanation of double
$J/\psi$ hadroproduction and thus gives strong phenomenological
support for the presence of intrinsic heavy quark states in hadrons.
While the general agreement with the intrinsic charm model is quite
good, the excess events at medium $x_{\psi \psi}$ suggests that intrinsic
charm may not be
the only $\psi \psi$ QCD production mechanism present or that the
model parameterization with a
constant vertex function is too
oversimplified.  The $x_{\psi \psi}$ distributions can also be affected by the
$A$ dependence.  Additional mechanisms, including an update of
previous models \cite{BHK,ES,HM,Russ,LL}, will be presented in
a separate paper \cite{RV}.

The intrinsic heavy quark
model can also be used to predict the features of heavier
quarkonium hadroproduction, such as $\Upsilon \Upsilon$,
$\Upsilon \psi $, and $(\bar b c)(\bar c b)$ pairs.
Using $\widehat{m}_b = 4.6$ GeV, we find
that the single $\Upsilon$ and $\Upsilon \Upsilon$ pair $x$
distributions are similar to the equivalent $\psi \psi$
distributions.  The average mass, $\langle M_{\Upsilon \Upsilon}
\rangle$, is 21.4 GeV for pion projectiles and 21.7 GeV for a
proton, a few GeV above threshold, $2m_\Upsilon = 18.9$ GeV.  The
$x_{\Upsilon \psi}$ pair distributions are also similar to the $\psi
\psi$ distributions but we note that $\langle x_\Upsilon \rangle =
0.44$ and $\langle x_\psi \rangle = 0.30$ from a $\vert \overline u
d c \overline c b \overline b \rangle$ configuration and $\langle
x_\Upsilon \rangle = 0.39$ and $\langle x_\psi \rangle = 0.27$ from
a $\vert u u d c \overline c b \overline b \rangle$ configuration.
Here $\langle M_{\Upsilon \psi} \rangle = 14.9$ GeV with a pion
projectile and 15.2 GeV with a proton, again a few GeV above
threshold, $m_\Upsilon + m_\psi = 12.6$ GeV.

It is clearly important for
the double $J/\psi$  measurements to
be repeated with
higher statistics and also at higher energies.
The same intrinsic Fock states will also
lead to the production of multi-charmed baryons in the
proton fragmentation region.
It is also interesting to study the correlations of the heavy
quarkonium pairs to search for possible new four-quark bound states
and final state interactions generated by multiple gluon exchange
\cite{LL}. It has been suggested that such QCD Van der Waals
interactions could be anomalously strong at low relative rapidity
\cite{Manohar,Bdet}.

There are many ways in which the intrinsic heavy quark
content of light hadrons can be tested.
More measurements of the charm and bottom structure
functions at large  $x_F$ are needed to confirm
the EMC data \cite{EMCic}.  Charm production in the
proton fragmentation region in deep inelastic
lepton-proton scattering is
sensitive to the hidden charm in the
proton wavefunction. The presence of intrinsic heavy
quarks in the hadron wavefunction
also enhances heavy flavor production in hadronic
interactions near threshold.
More generally, the intrinsic heavy quark model
leads to enhanced open and hidden heavy quark production
and leading particle correlations at high $x_F$ in hadron
collisions with a distinctive strongly-shadowed
nuclear dependence characteristic of soft hadronic collisions.

\vspace{0.5cm}
\begin{center}
{\bf Acknowledgements}
\end{center}

We thank J. Grunhaus, P. Hoyer, G. Ingelman, J.-C. Peng,  W.-K. Tang,
and M. V\"anttinen for
helpful discussions.

\newpage
\begin{center}
{\bf Figure Captions}
\end{center}
\vspace{0.5in}

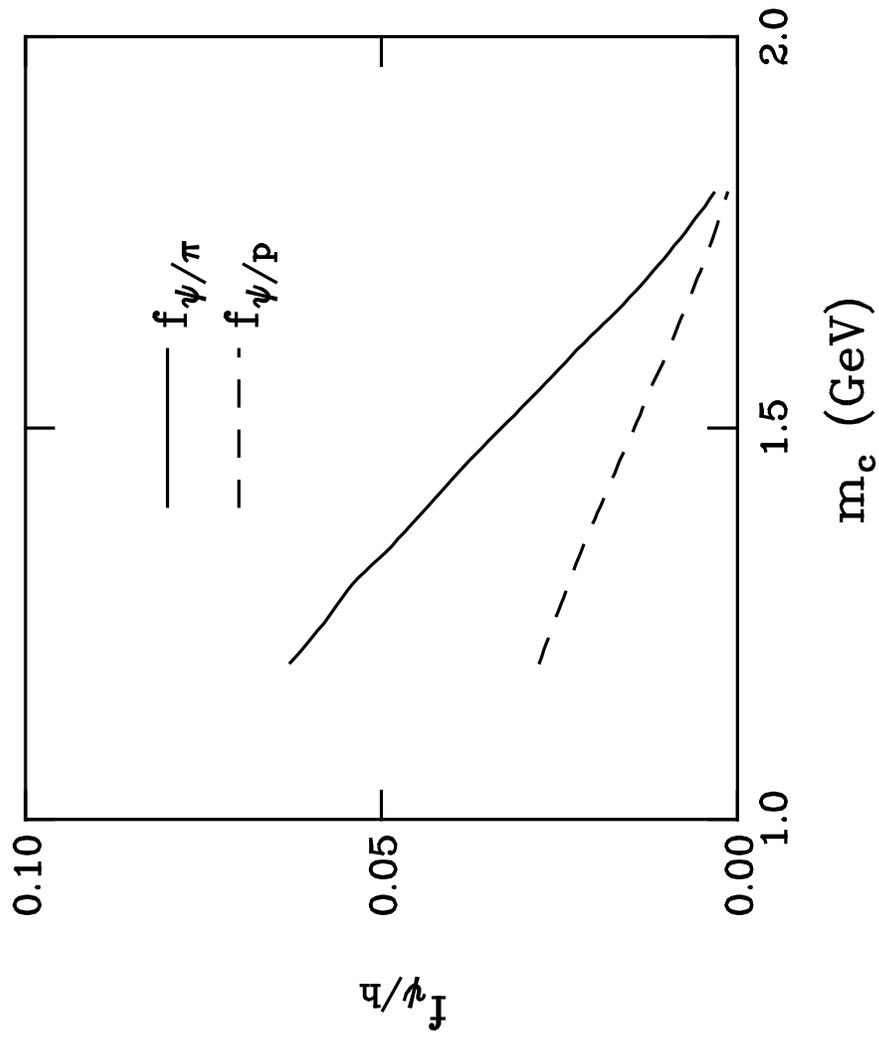
\noindent Figure 1.
The fraction of intrinsic $c \overline c$ states below the $D
\overline D$ threshold that produce $J/\psi$'s is given as a function
of charmed quark mass for pion
(solid) and proton (dashed) projectiles. \vspace{0.2in}

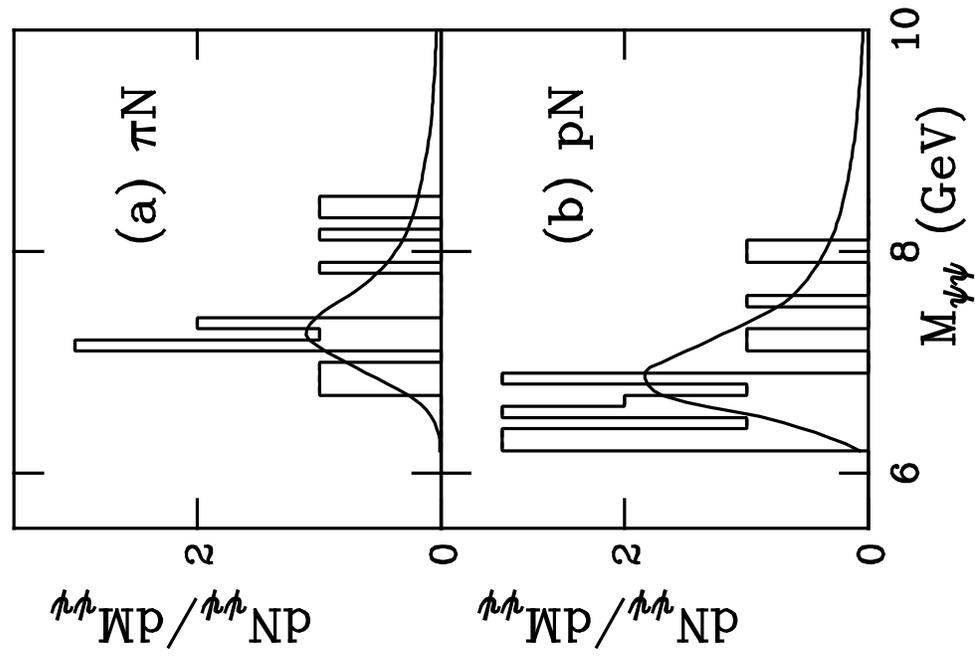
\noindent Figure 2.
The calculated $\psi \psi$ mass distributions, Eq.\ (8), are
compared with histograms of the NA3 data \cite{Badpi,Badp} from
$\pi^- N$ (a) and $pN$ (b) production.  The calculations are
normalized to the total number of $\psi \psi$ events. \vspace{0.2in}

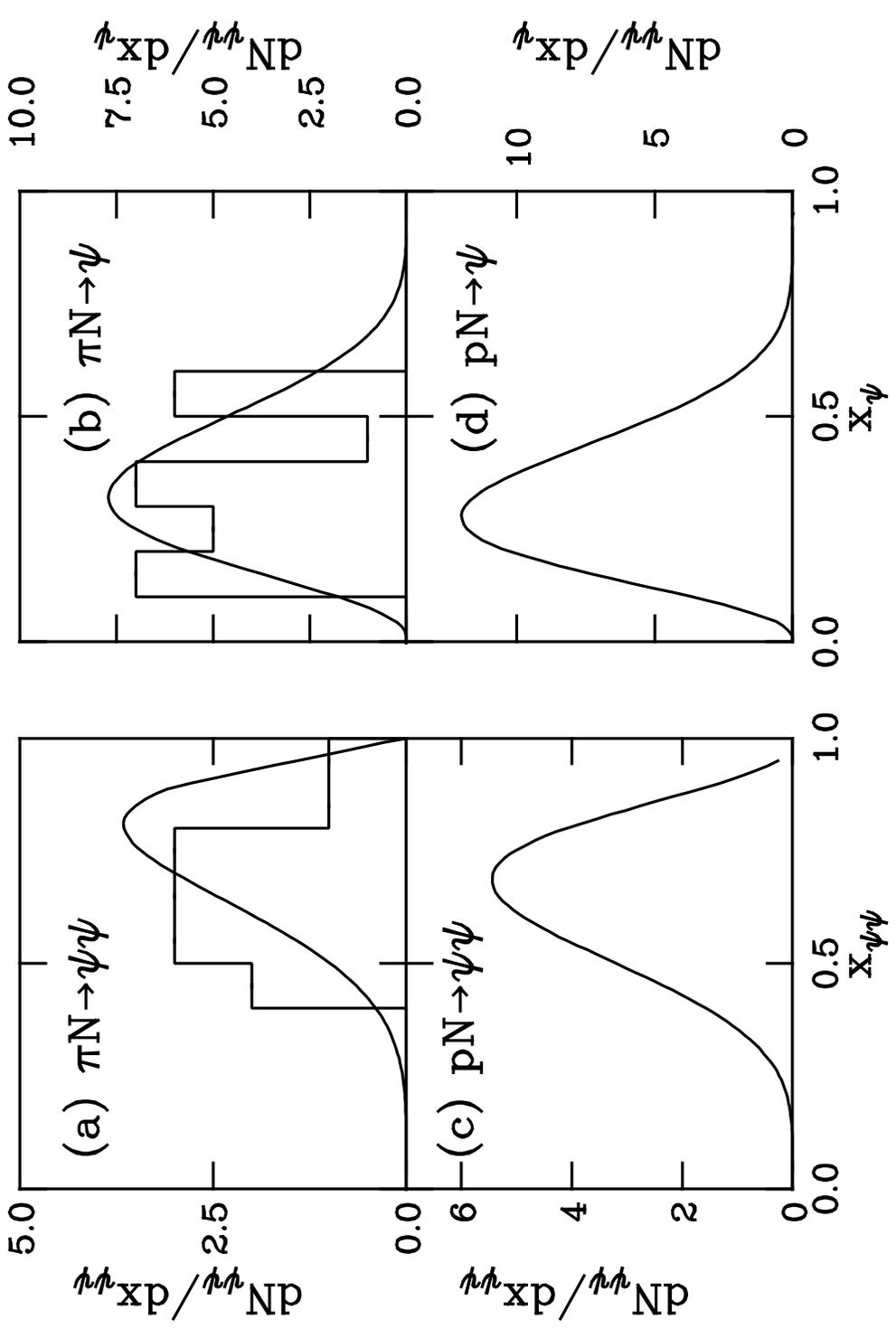
\noindent Figure 3.
The $\psi \psi$ pair distributions are shown in (a) and (c) for the
pion and proton projectiles.  Similarly, the distributions of
$J/\psi$'s from the pairs are shown in (b) and (d).  Our
calculations are compared with the $\pi^- N$ data at 150 and 280
GeV$/c$ \cite{Badpi}.  The $x_{\psi \psi}$ distributions are
normalized to the number of pairs from both pion beams (a) and the
number of pairs from the 400 GeV proton measurement (c). The number
of single $J/\psi$'s is twice the number of pairs.

\end{document}